\documentclass[sigconf]{acmart}
\setcopyright{none}
\renewcommand\footnotetextcopyrightpermission[1]{}
\usepackage{setspace}
\usepackage{setspace}
\usepackage{booktabs} 
\usepackage{soul,xcolor}
\usepackage{listings}
\usepackage{amsmath}
\usepackage{color}
\usepackage{xcolor}
\usepackage{lipsum}
\usepackage{comment}
\usepackage[title]{appendix}
\usepackage{multirow}
\usepackage{booktabs}
\usepackage{xspace}
\usepackage{graphicx}
\usepackage{caption}
\usepackage{pdfpages}
\usepackage{setspace}
\usepackage{subfiles}
\usepackage{booktabs}
\usepackage{array}
\usepackage{subcaption}
\usepackage[utf8x]{inputenc}
\usepackage[font=small,labelfont=bf]{caption}
\usepackage{url}
\setcopyright{rightsretained}
\usepackage{enumitem}
\setitemize{noitemsep,topsep=0pt,parsep=0pt,partopsep=0pt}
\RequirePackage{filecontents}
\newcolumntype{P}[1]{>{\centering\arraybackslash}p{#1}}

\newcommand{\startcompact}[1]{\par\vspace{-0.75em}\begin{#1}%
\allowdisplaybreaks\ignorespaces}

\newcommand{\stopcompact}[1]{\end{#1}\ignorespaces}

\newcommand{\name}{EasyCrash\xspace}
\newcommand{\tname}{NVCT\xspace}
\definecolor{dong}{RGB}{0,0,255}
\definecolor{dong_comment}{RGB}{0,255,0}

\begin{document}
\title{\name: Exploring Non-Volatility of Non-Volatile Memory for High Performance Computing Under Failures}
\author{Jie Ren}
\affiliation{
  \institution{University of California, Merced}
}
\email{jren6@ucmerced.edu}

\author{Kai Wu}
\affiliation{
  \institution{University of California, Merced}
}
\email{kwu42@ucmerced.edu}

\author{Dong Li}
\affiliation{
  \institution{University of California, Merced}
 }
 \email{dli35@ucmerced.edu}

\begin{abstract}
Emerging non-volatile memory (NVM) is promising for building future HPC. Leveraging the non-volatility of NVM as main memory, we can restart the application using data objects remaining on NVM when the application crashes. This paper explores this solution to handle HPC under failures, based on the observation that many HPC applications have good enough intrinsic fault tolerance. To improve the possibility of successful recomputation with correct outcomes and ignorable performance loss, we introduce \name, a framework to decide how to selectively persist application data objects during application execution. Our evaluation shows that \name transforms 54\% of crashes that cannot correctly recompute into the correct computation while incurring a negligible performance overhead (1.5\% on average). Using \name and application intrinsic fault tolerance, 82\% of crashes can successfully recompute. When \name is used with a traditional checkpoint scheme, it enables up to 24\% improvement (15\% on average) in system efficiency. 

\end{abstract}

\maketitle

\section{Introduction }
The extreme-scale high performance computing (HPC) systems face a grand challenge on system reliability. Transient hardware faults 
are one of the major concerns of current HPC systems~\cite{Snir:2014:AFE:2747699.2747701}. 
Those faults often result in application
failures (application crashes). 
Application crashes lose the application's work and decrease HPC system efficiency. 
A typical HPC system nowadays has a mean-time between failure (MTBF) of tens of hours \cite{1559953,Egwutuoha2013,4629264,inproceedings},
even with hardware- and software-based protection mechanisms. It is expected that the failure rate could further increase in the future, as 
the complexity of the HPC systems increases. This indicates that a larger portion of computation cycles will have to be used to handle application failures~\cite{whitepaper,Gupta:2017:FLS:3126908.3126937}. 

Byte-addressable non-volatile memory (NVM) technologies, such as 
Intel Optane DC persistent memory DIMM~\cite{Intel3D}, are emerging. 
NVM can provide better density and energy efficiency than DRAM while providing DRAM-like performance. 
Recent efforts have demonstrated the possibility of using NVM as main memory~\cite{Mnemosyne:ASPLOS11,nvtree:fast15,Kolli:ASPLOS2016,loopbased:pact17, Pelley:isca14} with \texttt{load/store} instructions and for future HPC~\cite{unimem:sc17,shuo:cluster17, nvm_ipdps2012,Fernando:2018:NAH:3208040.3208061,6569798, sc18:wu}. 
In this paper, we leverage non-volatility of NVM as main memory, and explore a novel solution to handle HPC applications under failures, aiming to improve system efficiency. 

One way to leverage the non-volatility of NVM for HPC under failures is to use NVM as a fast persistent media to implement the traditional checkpoint/restart (C/R). C/R is the most common fault tolerance mechanism in HPC. C/R periodically saves application data (a checkpoint) into persistent media. Once a failure
happens, C/R restarts the application~\cite{1592865,5695644} by loading a previously
saved intermediate state of the application (i.e., a checkpoint). Given high bandwidth of NVM (comparing with traditional hard drive), we can save application data into local NVM of each node, which reduces checkpoint overhead and improves system efficiency.

However, using NVM to build C/R 
has limitations. 
First, creating checkpoints in NVM (used as main memory) can double or even triple memory footprint of the application, and hence reduces the effective capacity of NVM. This is especially problematic for those scientific simulations with large data sets. For those applications, reducing the effective capacity of NVM constrains the simulation scale that the scientists can study. Second, it worsens the endurance problem faced by NVM. NVM has limited endurance and can tolerate a limited number of writes. 
For example, the write endurance of phase change memory (a promising NVM technology) is seven orders of magnitude lower than DRAM~\cite{5375309}. 
As a result, the endurance problem of NVM (used as main memory) have been actively studied recently~\cite{8573121, 8552412,  pldi13:gao, Lee:2009:APC:1555754.1555758, lazypersistency:isca18,5430747}. Since checkpointing must be written to persistent NVM, checkpointing can cause a number of additional writes in NVM.
In this paper, we introduce an application-level solution (named  \textit{\name}) that explores the non-volatility of NVM to handle application failures. \name does not create data copy as C/R does. Instead, it flushes cache blocks of some data objects of the application to persist them at certain execution phases of the application. 
When the application crashes, data objects in NVM are not lost (although some updates in caches are lost), and the application restarts using remaining data objects in NVM. Using \name, we aim to improve HPC system efficiency by reducing the frequency of checkpoint and reduce writes to NVM.

The design of \name is based on three observations. 
First, many HPC applications are characterized with large data sets and most of them may not be in caches during application execution, because of limited cache capacity. This indicates that using cache flushing (instead of making data copy) to persist data objects can potentially save a large number of writes. 

Second,  \name brings a challenge on data consistency when a crash happens, which can impact application recomputability. However, some HPC applications have intrinsic tolerance to data inaccuracy, which can be leveraged to tolerate data inconsistency. In particular, \name only ensures data consistency (between caches and main memory) right after cache flushing. When a random crash happens, a data object in NVM and caches may not be consistent, because of out-of-order stores in NVM and writing-back caching. 
Applications may not recompute successfully after recovering from a crash.
However, some HPC applications, such as iterative solvers (e.g., the preconditioned conjugate gradient method, Newton method, and multigrid method), Monte Carlo-based simulations~\cite{xsbench2:john} and some machine learning workloads (e.g., Kmeans and CNN training), have natural error resilience to localized numerical perturbations, because they require computation results to converge over time. As a result, they can intrinsically tolerate some data inconsistency. Furthermore, flushing cache blocks at appropriate execution phases can reduce data inconsistency and improves application recomputability. 

Third, many HPC applications have application-specific acceptance verification (e.g., based on physical laws and math invariant). 
Leveraging the verification, the application can detect whether computation results are acceptable before delivering them to end users. For example, large-scale computational fluid dynamics simulations examine the correctness of the result by making the comparison to exact analytical results~\cite{Roache:580994}. 
Those applications with acceptance verification can reduce the probability of producing incorrect results that might be generated by applications.



%
\begin{figure}[!t]
    \centering
    \includegraphics[width=\linewidth]{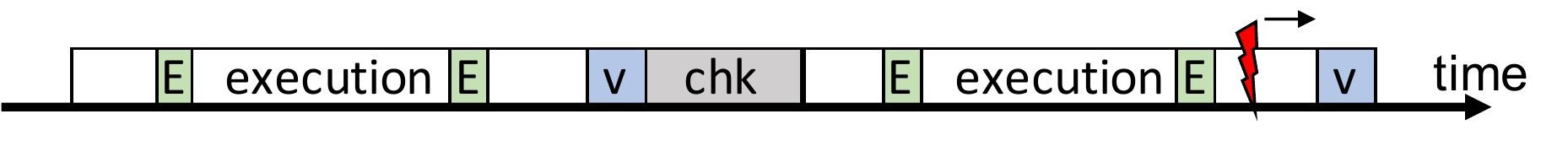} 
   
    \caption{ An illustration of how an HPC application behaves with \name. Figure annotation: ``chk'' - checkpoint; ``v'' - application acceptance verification; ``E'' - \name persistence operations.  }
	    \label{fig:EasyCrash}
\end{figure}
Figure~\ref{fig:EasyCrash} illustrates how \name works with NVM as main memory (such as Intel Optane DC persistent memory DIMM) for an HPC application. \name persists some data objects at certain execution phases of the application. Once a crash happens, the application immediately restarts using remaining consistent and inconsistent data objects in NVM. Application-specific acceptance verification 
checks if the recomputation result is correct. If the application cannot recompute successfully, then the application goes back to the last checkpoint. 
\name still needs C/R to guarantee 100\% of successful recomputation, but checkpointing can happen less frequently because \name makes some crashes successfully recompute without rolling back to the last checkpoint. Reducing the checkpoint frequency is very 
useful to improve system efficiency. 
It was reported that up to 50\% time in HPC data centers is spent in checkpointing~\cite{6258034,Ian:2005}.

Comparing with the traditional C/R, \name has the following benefits. 
First, \name does not create data copy, hence saving memory capacity and enabling scientific simulation with larger memory footprint. Second, \name flushes cache blocks using special instructions (e.g., \texttt{CLFLUSHOPT} or \texttt{CLWB}). Those instructions do not write back cache lines~\footnote{We distinguish cache line and cache block in the paper. The cache line is a location in the cache, and the cache block refers to the data that goes into a cache line.} to main memory, if the corresponding cache blocks are clean or not resident in caches; hence we reduce unnecessary writes to NVM and improve NVM lifetime. Third, writing to NVM is known to be expensive (comparing with writing to DRAM). Saving writes to NVM is beneficial for the performance of persisting data objects.

The most challenging design of \name is to decide how to persist data objects to ensure high recomputability without large runtime overhead. We employ analytical models and statistical analysis to address this challenge.
In particular, we characterize application recomputability using a number of crash tests. 
We use analytical models to decide where to persist data objects to ensure high recomputability and higher system efficiency than C/R without \name. Furthermore, to minimize runtime overhead of cache flushing, we use correlation analysis to decide which data objects are the most critical to application recomputability. \name only flushes cache blocks of those data objects. Such selective cache flushing reduces runtime overhead (also reducing the number of writes) while ensuring high recomputability.

To study application failures, we must have a tool that allows us to retain data objects in main memory for restart after a crash happens. The tool should faithfully reflect data inconsistency (between caches and main memory). 
The tool should also allow us to \textit{repeatedly} trigger crashes for study. The traditional systems (including hardware and software) cannot meet our needs: the volatile DRAM-based system loses data in main memory after a crash; the physical machines cannot tolerate repeated crash tests (tens of thousands of tests). The traditional random fault injection method~\cite{mg_ics12,bifit:SC12,Kai:icpp18} 
cannot work for us either, because there is no guarantee that each fault injection results in an application crash.

To address the above challenge, we introduce an emulation tool, named \textit{\tname} (standing for \textit{N}on-\textit{V}olatile memory \textit{C}rash \textit{T}ester).
In essence, the tool is a PIN~\cite{Pin:PLDI05}-based cache simulator plus rich functionality for crash tests. 
The tool allows the user to trigger application crash randomly, and then perform postmortem analysis on data values in caches and memory. 

To best of our knowledge, \name is the first comprehensive study that explores the non-volatility of NVM for HPC under failures. Existing work that focuses on crash consistency in NVM~\cite{ Dulloor:eurosys14, nvheap:asplos11, Kolli:ASPLOS2016, Mnemosyne:ASPLOS11,cdds:fast11} enforces data persistency for enterprise workloads with strong requirements on transaction semantics, which brings large runtime overhead and infeasible for HPC~\cite{shuo:cluster17}. \name is a solution customized for HPC. 

In summary, this paper makes the following contributions:
\begin{itemize}[leftmargin=*]
    \item A novel method to leverage the non-volatility of NVM to restart HPC applications under failures; 
    \item An open-sourced tool~\footnote{
    The tool is available online. https://github.com/PASAUCMerced/NVC}
    to enable crash study on NVM; Based on our knowledge, this tool is the first one for such study.
    \item Characterization of application recomputability with different data objects persisted in NVM at different execution phases; 
    \item An evaluation of \name with a spectrum of HPC applications. \name is able to transform 54\% of crashes that cannot correctly recompute into the correct computation, while incurring a negligible performance overhead (1.5\% on average). Using \name and application intrinsic fault tolerance, 82\% of crashes can successfully recompute. As a result, when \name is used with C/R, it enables up to 24\% improvement (15\% on average) in system efficiency.
    Comparing with C/R without \name, \name reduces the number of writes by 44\% on average. 


    
\end{itemize}
\section{Background }
\subsection{Cache Flushing for Data Persistence}
\label{sec:cflash_data_persist}

Because of the prevalence of volatile caches, data objects in applications may not be persistent in NVM when a crash happens. To ensure persistency and consistency of data objects in NVM, the programmer typically employs ISA-specific cache flushing instructions (e.g., \texttt{CLFLUSH}, \texttt{CLFLUSHOPT}, and \texttt{CLWB}). 

To persist a large data object, the current common practice is to flush all cache blocks of the data object~\cite{pmdk}, even when some of them are not in the cache. This is because we do not have a mechanism to track dirty cache lines and whether a specific cache block is resident in the cache. However, flushing a clean cache block or a non-resident cache block is much less expensive than flushing a dirty one resident in the cache, because there is no writeback. 

\subsection{Terminology and Problem Definition}

\textbf{Data objects.}
We focus on heap and global data objects in this paper, and do not consider stack data objects. 
Making such a selection on data objects is based on our comprehensive survey on 51 HPC applications: in our survey, we find that major memory footprint and most important data objects (important to execution correctness) in HPC applications are heap and global ones. Our observation is aligned with the recent work~\cite{Ji:2017:UOM:3126908.3126917, bifit:SC12}. 

We study data objects (but not the whole system state) for recomputation study, because of two reasons: (1) the current main-stream NVM programming models for NVM~\cite{pmdk, nvheap:asplos11,Mnemosyne:ASPLOS11} focus on persisting data objects for the convenience of application restart; (2) persisting the whole system state can cause large performance overhead. 

\textbf{Application recomputability.}
We define application recomputability in terms of application outcome correctness. In particular, we claim an application recomputes successfully after a crash, if the final application outcome remains correct. The application outcome is deemed correct, as long as it is acceptable according to application semantics. Depending on application semantics, the outcome correctness can refer to precise numerical integrity (e.g., the outcome of a multiplication operation must be numerically precise), or refer to satisfying a minimum fidelity threshold (e.g., the outcome of an iterative solver must meet certain convergence thresholds). Leveraging application-level acceptance verification, we can determine the correctness of the application outcome. 

Furthermore, we define application recomputability with a high requirement on performance to make our solution practical for HPC. In particular, for an HPC application with iterative structures, we claim that it recomputes successfully when its outcome is correct \textit{and} it does not take extra iterations to finish.

Application recomputability quantifies the possibility that once a crash happens, the application recomputes successfully. To calculate application recomputability, one has to perform a number of crash tests to ensure statistical significance. Each test triggers a random crash and restarts the application. We use the ratio of the number of tests that successfully recompute to total number of tests as the \textit{application recomputability}. We call all of the crash tests as a \textit{crash test campaign}.  


We distinguish ``restart'' and ``recompute''. After the application crashes, the application may resume execution, which we call \textit{restart}. If the application outcome is correct and there is no need of extra iterations to finish, we claim the application  \textit{recomputes.} 

\textbf{System efficiency.} It is defined as the ratio of the accumulated useful computation time to total time spent on the HPC system. The total time includes useful computation time, checkpoint time, lost computation time because of crashes, and recovery time.

\textbf{Application target.} We focus on HPC applications. The effectiveness of \name is affected by the acceptance verification and resilience characteristics of those applications. 

The acceptance verification can happen at the end of the application~\cite{hpl} or during the application execution~\cite{nicholaeff2012cell}. The acceptance verification detects whether the application state is corrupted before delivering results to end users. Typically it is the programmer's responsibility to write the acceptance verification to ensure that computation results do not violate application-specific properties (e.g., convergence conditions or numeric tolerance for result approximation). The application-level acceptance verification is very common in HPC applications, and increasingly common, because of the strong needs of increasing the confidence in the results offered by HPC applications. These needs are driven by increasing awareness of hardware faults and application complexity.  


A large class of HPC applications with iterative structures are naturally resilient to computation inaccuracy~\cite{mg_ics12, Chippa:2013:ACI:2463209.2488873} and often characterized with a main computation loop dominating computation time. 
Those iterative applications are promising to be recomputable after crashes because they work by improving the accuracy of the solution step by step, and the process can help to eliminate errors. 
For example, a convergent iterative method can tolerate inconsistent data during the convergence process. Because of the prevalence and importance of those applications, the recent work on approximate computing also focuses on those applications~\cite{ics06:rinard, 19-mengte2010exploiting, oopsla13:carbin, oopsla14:misailovic, onward10:rinard, fse11:douskos}.

\textbf{Failure model.} 
We focus on application failures which could be caused by power loss or processor failures. We do not consider application failures caused by software bugs, because those bugs can prevent application recomputation.
\subsection{Optane DC Persistent Memory Module}
The very recent release of Intel Optane DC persistent memory module (PMM) is arguably the most mature NVM product as \textit{main memory} and promising for future HPC~\cite{aurora_anl}. We put our discussion in the context of this hardware to make our work more useful.

Optane DC PMM can be configured as either memory mode or app-direct mode. With the memory mode, Optane DC PMM does not provide data persistency, hence not relevant to our work. We assume that Optane DC PMM uses \textit{app-direct} mode in our work. With this mode, Optane DC PMM is provisioned as persistent memory with byte addressability. The application can directly access it using load/store instructions without going through the DRAM cache, 
and flushing CPU caches makes data persistent in Optane DC PMM. To locate data objects in Optane DC PMM after a failure, the user can leverage a memory-mapped file-based mechanism. This mechanism uses address offset, which is independent of memory addresses, to keep track of the location of data objects. This mechanism is commonly used in the exiting NVM-aware programming models~\cite{intel_nvm_lib, mnemosyne_asplos11, usenix13:rudoff}.
\section{\tname: A Tool for Studying Application Recomputability}
\label{sec:NVCT}
To enable our study on application recomputability in NVM, we introduce a PIN-based crash emulator, \tname. \tname includes a simulated multi-level, coherent cache hierarchy and main memory, a random crash generator, a set of APIs to support the configuration of crash tests and application restart, and a tool to examine data inconsistency for post-crash analysis. Different from the traditional PIN-based cache simulator, \tname not only captures  microarchitecture level, cache-related hardware events such as cache misses and invalidation, but also records the most recent values of data objects in the simulated caches and main memory. 

\definecolor{codegreen}{rgb}{0,0.6,0}
\definecolor{codegray}{rgb}{0.5,0.5,0.5}
\definecolor{codepurple}{rgb}{0.58,0,0.82}
\definecolor{backcolour}{rgb}{0.95,0.95,0.92}
\lstdefinestyle{style1}{
    commentstyle=\color{codegreen},
    keywordstyle=\color{magenta},
    numberstyle=\tiny\color{codegray},
    stringstyle=\color{codepurple},
    basicstyle=\footnotesize,
    frame=bottomline,
    numbers=left,                    
    numbersep=5pt, 
    keywordstyle=\color{red},
    emph={write_back_invalidate_cache,cache_line_write_back}, 
    emphstyle=\color{blue},
	escapeinside={(*@}{@*)},
}
\lstset{style=style1}

\begin{figure}[!t]
 \centering
    \begin{subfigure}[b]{\linewidth}
\begin{lstlisting}[language=c, 
xleftmargin=.02\textwidth, xrightmargin=0.01\textwidth]
static double u[NR];
static double r[NR];
void main(int argc, char **argv) {
  int it;
  initialize();
  for (it = 1; it <= nit; it++) {//main computation loop
    for () { // a first-level inner loop; R1
        ...
        for() {...}// a second-level inner loop
    }
    for () { // a first-level inner loop; R2
        ...
    }
    for () { // a first-level inner loop; R3
        ...
    }
    for () { // a first-level inner loop; R4
        ...
    }
    cache_block_flush(u, NR*sizeof(double));
    cache_block_flush(r, NR*sizeof(double));
    cache_block_flush(&it, sizeof(int));
  }
  //result verification
  ...
}
\end{lstlisting}
\subcaption[a]{Persisting data objects during the application execution. }
    \label{fig:flushing}
        \end{subfigure}
\hfill 
\begin{subfigure}[b]{\linewidth}
\begin{lstlisting}[language=c, 
xleftmargin=.02\textwidth, xrightmargin=0.01\textwidth]
static double u[NR];
static double r[NR];
void main(int argc, char **argv) {
    int it,it_old;
    initialize();
    load_value(u,NR*sizeof(double));
    load_value(r,NR*sizeof(double));
    load_value(&it_old,sizeof(int));
    for (it = it_old; it <= nit; it++) {//main comp loop
  	...
  	//flush cache blocks
    }
    //result verification
    ...
}
\end{lstlisting}
\subcaption[b]{Restart MG. }
\label{fig:restart}
\end{subfigure}
\caption{Study recomputability of MG with \tname.} 
\label{fig:example_app_restart}	   
\end{figure}

\textbf{Main memory simulation. }
Different from the microarchitectural simulation of main memory, the main memory simulation in \tname records data values and their corresponding memory addresses. Whenever the cache simulation writes back any cache line, the corresponding data values in the simulated main memory are updated. Using this method, we can easily determine data inconsistency between the caches and main memory. During a crash test, the data values of user-specified data objects in the simulated main memory can be dumped into a file for post-crash analysis. 

\textbf{Random crash generation.}  
\tname emulates the occurrence of a crash by stopping application execution after a randomly selected instruction. 
Furthermore, \tname can report call path information when a crash happens. This is implemented by integrating CCTLib~\cite{Chabbi:2014:CPP:2581122.2544164} into \tname.
The call path information introduces program context information for analyzing crash results. The context information can help to distinguish those crash tests that happen in the same program statement, but with different call stacks. 

\textbf{Calculation of data inconsistent rate.} 
\tname reports data inconsistent rate after a crash happens. 
When a crash happens, \tname examines dirty cache lines in the simulated cache hierarchy. 
\tname compares each dirty cache line with the corresponding cache block in main memory to determine the number of dirty data bytes in the cache line. 
To calculate the data inconsistent rate for a data object, \tname examines the file dumped by the main memory simulation to determine the total number of dirty bytes in the data object, and then divide the number by the data object size. 

\textbf{Application restart}. 
When restarting, \tname loads data values from the file dumped by the main memory simulation to initialize user-specified data objects. Some data objects are initialized by the application itself. 
Then, \tname resumes the main loop, starting from the beginning of the iteration where the crash happens.

\textbf{Putting all together.} 
To use \tname, the user needs to insert APIs to specify (1) data objects that need to be persisted during application execution, 
(2) the initialization phase of the application for a restart, and (3) code regions where crashes can happen. The user also needs to configure cache simulation and crash tests (e.g., how many crash tests and what probability distribution the crash tests follow). During the application execution, \tname leverages the infrastructure of PIN to instrument the application and analyze instructions for cache and memory simulations. \tname triggers a crash as configured, and then performs post-crash analysis to report data inconsistent rate and restart the application. 

\textbf{An example}. Figure~\ref{fig:example_app_restart} gives an example of how we study application recomputability.
This is a multi-grid (MG) numerical kernel from the NAS parallel benchmark suite~\cite{nas} (NPB). Like many HPC applications, MG has a main computation loop, within which we persist two global data objects and a loop iterator~\footnote{In the rest of the paper, we always persist a loop iterator to bookmark where the crash happens. This makes restart easier. Persisting just one iterator has almost zero impact on application performance.} (Lines 20-22 in Figure~\ref{fig:flushing}) in each iteration. After a crash happens, we restart MG using Figure~\ref{fig:restart}. To restart, the application re-initializes computation (Line 5 in Figure~\ref{fig:restart}), loads the values of the two data objects and old loop iterator (Lines 6-8) from NVM, and restarts the main computation loop from the iteration where the crash happens (Line 9). We run MG to completion and verify the application outcome.
\section{Characterization of Application Recomputability}
We characterize application recomputability to motivate our design.
\subsection{Experiment Setup}
\label{sec:motivation_examples_setup}
\textbf{Benchmarks for evaluation.}
We use all benchmarks from NPB. To enrich our benchmark collection, we also choose botsspar from SPEC OMP 2012~\cite{10.1007/978-3-642-30961-8_17}, kmeans from Rodinia~\cite{5306797} and LULESH~\cite{LULESH:spec}. In total, we have 11 benchmarks, covering dense linear algebra, sparse linear algebra, spectral methods, structured-grid, graph traversal, and data mining. 
Those benchmarks are chosen, because of their representativeness and explicit code structures to verify application outcomes. For NPB benchmarks, we use Class C as input problems; For LULESH, botsspar and kmeans, we use 100, ref, and kdd.txt as the input problems respectively. We use these input problems, because the application memory footprints with them are larger than the last level cache size. 
Table~\ref{tab:bench} summarizes these benchmarks and shows their characteristics related to our study. 

\begin{table*}[ht]
\centering
\scriptsize
\caption{Benchmark information for crash experiments. ``R/W'' = ``Read/Write ratio'', ``DO'' = ``data object'', ``iter'' = ``iterations''. }
\label{tab:bench}
\begin{tabular}{|c|c|c|c|c|c|c|c|c|}
\hline
\textbf{Benchmarks} & \textbf{Description}   & \textbf{\begin{tabular}[c]{@{}l@{}}\# of code\\regions\end{tabular}} & \textbf{R/W} & \textbf{Memory footprint} &  \textbf{\begin{tabular}[c]{@{}l@{}}Candi. of critical \\ DO size\end{tabular}} & \textbf{\begin{tabular}[c]{@{}l@{}}Critical DO size\end{tabular}} & \textbf{\begin{tabular}[c]{@{}l@{}}Ave. \# of extra iter. to \\ restart (restart overhead)\end{tabular}}  
& \textbf{\begin{tabular}[c]{@{}l@{}}Total \# of iter. in the \\original app execution\end{tabular}} 
\\ \hline
\textbf{CG}                   & Sparse linear algebra  &        6                    &  21:1                   &    947MB                      &       5.7MB                      &   2.3MB     &9.1&75                       \\ \hline
\textbf{MG}                   & Structured grids       &      4                      &  7:1  &         3.4GB    &     2.3GB   &    1.2GB &0&20      \\ \hline
\textbf{FT}                   & Spectral method       &        4                  & 1:1      &   5.1GB       &       4.0GB     &       4.0GB  &0&20      \\ \hline
\textbf{IS}                   & Graph traversal (sorting)                &               8             &   2:1                  &           1.0GB    &    264MB                         &      4KB  &N/A(segfault) &10              \\ \hline
\textbf{BT}                   &  Dense linear algebra      &           15           &     2:1                &            644MB     &  525.6MB                           &           361.2MB &0&200   \\ \hline

\textbf{LU} & Dense linear algebra   &    4     &   5:2 &       644MB         &     599MB    &  164MB  &N/A (the verification fails)&250             \\ \hline

\textbf{SP}                   & Dense linear algebra       &   16                   &    2:1                 &    772MB               &           561MB    &   184MB             &0&400               \\ \hline

\textbf{EP} & Monte Carlo   &    2    &   2:1 &         1MB     &  1MB    &  80B  &N/A (the verification fails)&65535           \\ \hline

\textbf{botsspar}          & Sparse linear algebra   &         4                 &     2:1                &              3.74GB               &           3.36GB     &    3.36GB        &0&200            \\\hline
\textbf{LELUSH}               & Hydrodynamics modeling &                  4          &    5:1          &       567MB          &  251MB                           &     120MB &0&3517           \\ \hline
\textbf{kmean}        & Data mining  &  1   &   9:2                  &              222MB             &        20B      &     20B   &18.2&36 \\ \hline
\end{tabular}
\end{table*}

\textbf{System configuration.}
We simulate a three-level cache hierarchy (L1 cache:  32KB and 8-way set associativity; L2 cache: 1MB and 12-way set associativity; L3 cache: 19.25MB, 11-way set associative), with the 64B cache line, write-back, write-allocation, and LRU policies. This cache hierarchy is similar to that in Xeon Gold 6126. We use both single thread and multiple threads to run each benchmark. We show the results of single thread because of page limitation, but the conclusions we draw from the results of multiple threads are the same as those of single thread. 

\textbf{Crash tests.}
To ensure statistical significance, for each benchmark, we run a sufficient number of crash and recomputation tests (usually 1000-2000 tests), such that further increasing the number of tests does not cause big variation (less than 5\%) in the evaluation results. This method ensures that our evaluation is sufficient and our results are statistically correct. During application execution, we randomly stop it for crash tests. The times when the execution is stopped follow 
a discrete uniform distribution. This method of interrupting applications is common in the research on system fault tolerance~\cite{Kai:icpp18, bifit:SC12, sc18:guo,guan2014f,mg_ics12}. 
\subsection{Experiment Results}
\begin{figure}
\centering
\includegraphics[width=0.48\textwidth]{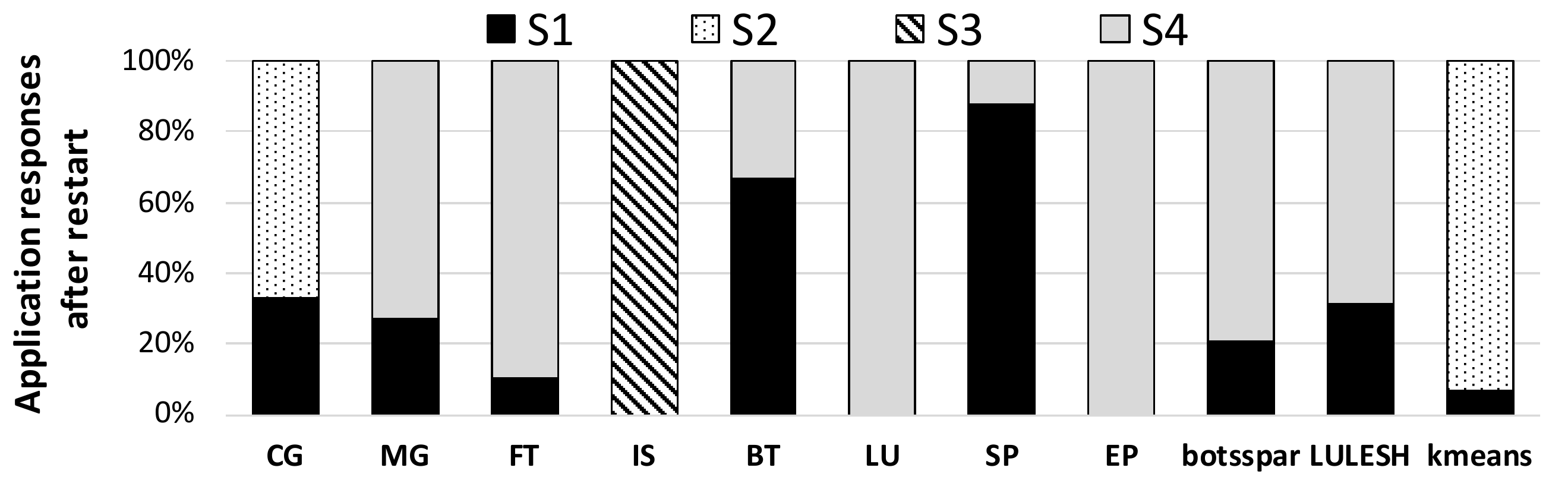}
\caption{Application responses after crash and restart. Figure annotation: S1 - successful recomputation without using extra iterations, S2 - successful recomputation with extra iterations, S3 - Interruption, and S4 - verification fails. }
\label{fig:recompur_success_rate}
\end{figure}

We observe four possible application responses after a crash and restart. (1) Successful recomputation without performance overhead: after a crash in the middle of an iteration and restart, the application successfully passes the acceptance verification, and uses no extra iteration to finish; 
(2) Successful recomputation with performance overhead: the application successfully passes the acceptance verification, but takes at least one more iterations than the original execution; (3) Interruption: the application cannot run to completion (e.g., due to segfault); (4) Verification fails: the application cannot pass the acceptance verification, even after taking two times as many iterations as in the original execution. 

Figure~\ref{fig:recompur_success_rate} and Table~\ref{tab:bench} (the last two columns) show the results based on the above classification. We notice that some applications show strong recomputability (e.g., 88\% for SP). 
Some applications (e.g., LU, IS, and EP) are the opposite: They cannot restart, or have segmentation faults.
We need to flush caches to improve their recomputability.


\textbf{Observation 1:} Different applications can have quite different recomputability.

\begin{figure}[htbp]
\centering
    \begin{subfigure}[b]{0.49\linewidth}
        \includegraphics[width=\textwidth]{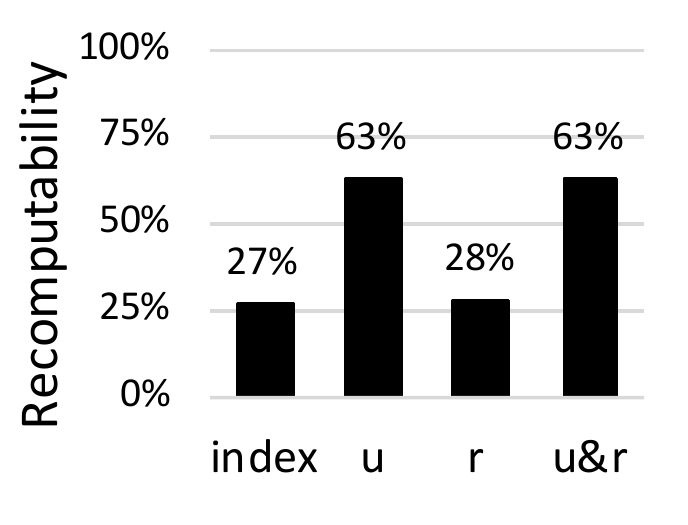}
        \vspace{-20pt}
        \subcaption[a]{}
    \label{fig:critical_data_obj}
        \end{subfigure}
    \begin{subfigure}[b]{0.49\linewidth}
    \includegraphics[width=\textwidth]{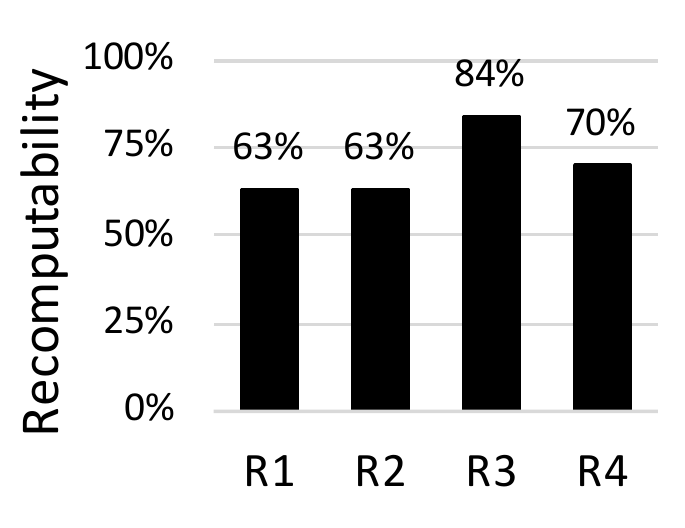}
    \vspace{-20pt}
            \subcaption[b]{}
        \label{fig:critical_code_region}
        \end{subfigure}
	    \caption{The recomputability of MG after (a) persisting three different data objects in MG; and (b) persisting $u$ in different code regions.} 
\label{fig:critical}
\end{figure}

To study how to improve application recomputability (i.e., the application passes the acceptance verification without using extra iterations), we selectively persist data objects and examine its impact on application recomputability. We do not persist all data objects, because that can cause large performance overhead (e.g., persisting all data objects in MG for just once causes 68\% performance overhead). 
Figure~\ref{fig:critical_data_obj} shows the results for MG. We choose three data objects ($index$, $u$ and $r$) for study. The three data objects are updated frequently and take a large portion of total memory footprint. We persist them at the end of each iteration of the main computation loop. Without persisting any data object, the recomputability of MG is 27\%.  By persisting the data object $u$, the recomputability is improved to 63\%; However, persisting the other two data objects, the recomputability is barely improved. 

\textbf{Observation 2:} Persisting different data objects have different implications on application recomputability. 

We further study the impact of where to persist data objects on application recomputability. MG has four first-level inner loops shown as R1-R4 in Figure~\ref{fig:flushing}. They represent four major execution phases. They all update $u$. 
We persist $u$ at the end of each execution phase. 
Figure~\ref{fig:critical_code_region} shows the result. 
Persisting $u$ at R3, we have 21\% improvement in recomputability, 
while persisting it at other code regions, we have smaller and similar (less than 7\%) improvement. 

\textbf{Observation 3:} The application has different recomputability, when persisting data objects at different code regions. 


The intuition explaining the observations 2 and 3 is that different data objects are associated with different memory access patterns and execution phases, and different code regions have different memory access patterns. As a result, data objects can show various inconsistent rates when a crash happens, and the success of application recomputation is sensitive to the data inconsistent rate.

\section{Design}
\label{sec:design}
Motivated by the above observations, we introduce \name, a framework that increases application recomputability with an ignorable runtime performance overhead and offers higher system efficiency than C/R without \name. \name \textit{automatically decides} which data objects should be persisted and where to persist them to maximize application recomputability and reduce writes in NVM. 
Based on the decision by \name, the user inserts APIs into the application to flush data objects exemplified in Lines 20-22 in Figure~\ref{fig:flushing}, which involves minor changes to the application. We describe how \name makes the decision in this section.

\subsection{Selection of Data Objects}
\label{sec:selection_data_obj}
We name data objects selected to be persisted, ``critical data objects'' in the rest of the paper.
To select data objects, we choose those data objects with the following properties as \textit{candidates}: (1) Their lifetime is the main computation loop; and (2) They are not read-only. Except for the candidates, the other data objects are either temporal or read-only, and not treated as the candidates of critical data objects. 
When the application restarts, the other data objects are not read from NVM. Instead, they are restored by either the initialization phase of the application or being re-computed based on the candidates of critical data objects. 
When the application restarts, the candidates are directly read from NVM. 

There is a large search space to select data objects out of the candidates. Assuming that there are $N$ candidates, then there are $2^N$ possible selections. We use statistical correlation analysis to select data objects efficiently.


Our selection method is based on the following observation. When a crash happens, data objects remaining in NVM can have different degrees of inconsistency. For example, a data object of 128MB could have 16MB of inconsistent data, giving an inconsistent rate of $16/128= 12.5\%$, while some data object could have an inconsistent rate of 50\%. We observe that the application recomputability correlates with inconsistent rates of some data objects, meaning that if these data objects have high inconsistent rate, the application recomputability is low. These data objects should be selected as critical data objects. We also observe that the application recomputability is not sensitive to the inconsistent rate of some data objects. Persisting these data objects does not matter to application recomputability. Hence, the sensitivity of application recomputability to the inconsistent rate of data objects can work as a metric to select data objects.

We use Spearman's rank correlation analysis~\cite{doi:10.1080/01621459.1972.10481251} to statistically quantify the correlation between the inconsistent rate of data objects and application recomputability. The result of the Spearman's rank correlation is the coefficient ($R_s$), which quantifies how well the relationship between two input vectors (data inconsistent rates and application recomputability) can be described using a monotonic function. Furthermore, we use the p-value of $R_s$ to ensure statistical significance of our analysis. The p-value is the probability of observing data that would show the same correlation coefficient in the absence of any real relationship between the input variables. 


To use the Spearman's rank correlation analysis, we build two vectors for each candidate data object, using the results collected from a  crash test campaign: One vector is composed of data inconsistent rates; The other vector is composed of application recomputation results (i.e., whether the application recomputes successfully or not). Each component of the two vectors is collected in one crash test. The vectors are used as input to the correlation analysis. 

Based on the Spearman's rank correlation analysis, we use two criteria to select data objects. (1) A critical data object should have a negative value of the correlation coefficient which indicates decreasing data inconsistent rate improves application recomputability. (2) The p-value of $R_s$ should be smaller than a threshold. We use 0.01 as the threshold, because it is a common threshold~\cite{doi:10.1080/01621459.1972.10481251}, and less than it statistically shows a very strong correlation in our study. 

\begin{figure}
\centering
\includegraphics[width=0.48\textwidth]{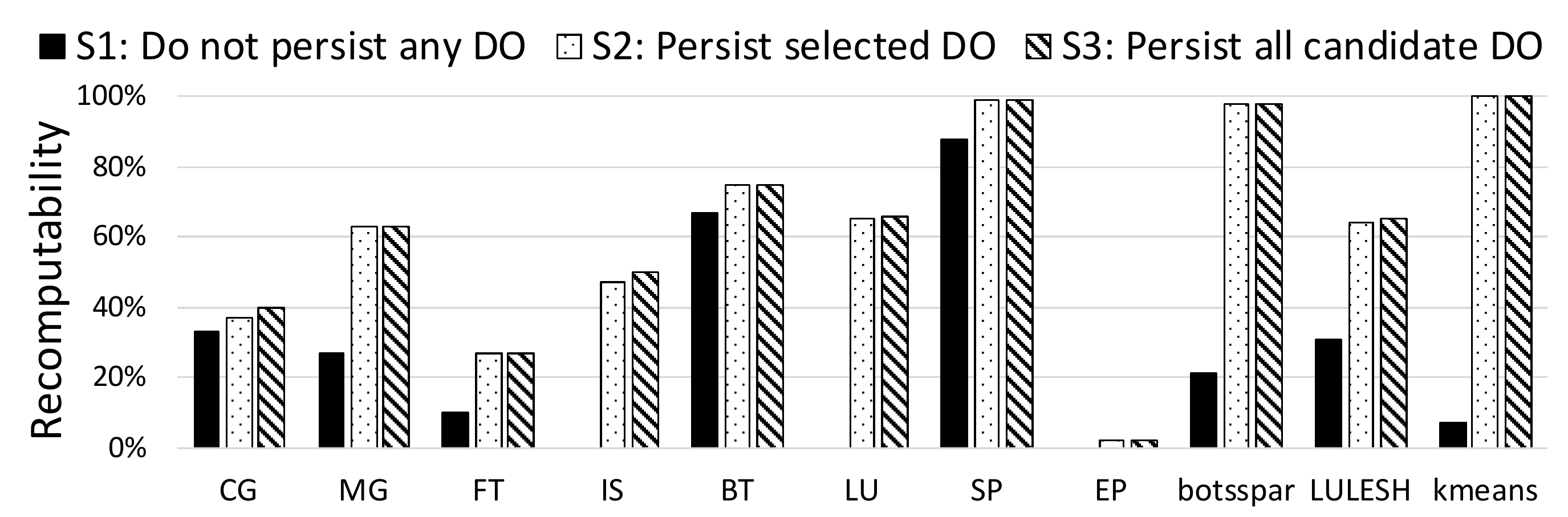}
\caption{Application recomputability under three strategies to persist data objects. Figure annotation: ``DO'' stands for data objects.} 
\label{fig:persistent_critical_data_object}
\end{figure}

\textbf{Verification of the selection of data objects.} To verify that our selection is effective, we evaluate application recomputability with three strategies: (1) Do not persist any data object; (2) Persist selected data objects; (3) Persist all candidate data objects. 
Figure~\ref{fig:persistent_critical_data_object} shows the results. The figure shows that the difference in application recomputability between (2) and (3) is less than 3\% in all cases. This verifies the effectiveness of our selection of data objects. 


\subsection{Selection of Code Regions}
\label{sec:selection_code_region}


In this section, we first define what are code regions in a typical HPC application. Then we formalize our problem of selecting code regions, and introduce an algorithm to solve it. Table~\ref{tab:annotation} summarizes the annotation for our formulation. 

\textbf{Application code regions.}
We characterize HPC applications as a set of iterative structures or loops. In particular, there is usually a main computation loop in an HPC application. Within the main loop, there are a number of inner loops that are typically used to update data objects iteratively. Figure~\ref{fig:flushing} shows an example of such a program abstraction for MG.

We model an application as a chain of code regions delineated by loop structures. A code region is either a first-level inner loop or a block of code between two adjacent, first-level inner loops. We use the above definition of code regions, because such code regions easily represent computation phases of the application. Persisting data objects in each code region ensures that the most recent computation results in a phase are persistent in NVM, and can effectively improve application recomputability. The similar definition of code regions can be found in~\cite{sc18:guo} to study application resilience to errors.

\begin{table}[!tb]
\small
\centering
\caption{Model annotation} 
\label{tab:annotation}
\begin{tabular}{|p{1.3cm}|p{6.4 cm}|}
\hline
\textbf{Parameter} & \textbf{Description}                                                                                            \\ \hline
\textbf{$R_s$}      & Spearman's rank correlation coefficient                                                                         \\ \hline
\textbf{$a_k$}      & The ratio of the accumulated execution time of the code region $k$ to the total execution time of the application \\ \hline
\textbf{$c_k$}      & Recomputability of the code region $k$                                                                               \\ \hline
\textbf{$Y$}        & Application recomputability                                                                                    \\ \hline
\textbf{$l_k$}      & The performance loss due to persistence operations in the code region $k$                                           \\ \hline
\textbf{$t_s$}      & Runtime overhead because of persistence operations in the application                                                             \\ \hline
\textbf{$\tau$}     & The system efficiency goal for long-running applications with \name                              \\ \hline
\textbf{$N$ and $M$} & \# checkpoint and \# crashes in the whole system time 
                        \\ \hline
\textbf{$W$}          & \# code regions in the application 
                        \\ \hline
\textbf{$M'$}        & \# crashes that go to the last checkpoint for recovery after using \name  
                        \\ \hline
\textbf{$M''$}       & \# crashes that use EasyCrash to recompute successfully
                        \\ \hline
\textbf{Any para. with ``prime''}  &   The corresponding parameters after applying \name
                        \\  \hline
\textbf{$c_k^{max}$}    &   Maximum recomputability of the code region $k$ after persisting data objects 
                        \\  \hline
\textbf{$x$}    &   The frequency to persist data objects in a loop-based code region
                        \\  \hline      
\textbf{$c_k^{x}$}    &   Recomputability when using $x$ as the frequency for cache flushing in the code region $k$ 
                        \\  \hline            
\end{tabular}
\end{table}

\textbf{Problem formulation.}
Among all code regions, we want to select code regions to satisfy two performance goals. (1) \textit{The runtime performance goal}: the application with critical data objects persisted at the selected code regions should have runtime overhead smaller than a threshold $t_s$. $t_s$ is set by the user ($t_s = 3\%$ of the application execution time without any crash in our study). (2) \textit{The system efficiency goal} for long-running applications: the system efficiency with \name (including successful and unsuccessful recomputation) should be better than that with C/R without \name. 
Achieving this goal requires that the recomputability of the application with the selected code regions should be higher enough (higher than a threshold $\tau$). Section~\ref{sec:checkpoint_emulation} discusses how to decide $\tau$.  

We name the selected code regions as ``critical code regions'' in the rest of the paper. In the following discussion, we assume that there is only one critical code region, in order to make our formalization easy to understand. But our formalization can be easily extended to multiple critical code regions.

Assume that there are $W$ code regions in an application and $Y$ is the application recomputability without persisting any data objects. The recomputability of a code region $i$ is $c_i$. \textit{The recomputability of a code region} is the possibility that when a crash happens during the execution of the code region, the application can be  successfully recomputed. We formulate $Y$ as follows, based on the definition of recomputability.
\begin{equation}
    Y = \sum_{i=1}^{W} (a_i \times c_i)
\end{equation}

\noindent where $a_i$  is the ratio of the accumulated execution time of the code region $i$ to the total execution time of the application. 

Assume that the code region $k$ ($1 \le  k \le W$) is selected as a critical code region. After persisting critical data objects at the code region $k$, the recomputability of the application and code region becomes $Y^{\prime}$ and $c_k^{\prime}$ respectively. 
We have performance loss $l_k$ for persisting critical data objects in the code region $k$, which is the ratio of the absolute performance loss to the total execution time.

$Y^{\prime}$ can be calculated based on $c_k^{\prime}$, shown in Equation~\ref{eq:new_Y}.
\begin{equation}
\label{eq:new_Y}
    Y^{\prime} = \sum_{i=1}^{k-1} (a_i^{\prime} \times c_i) + a_k^{\prime} \times c_k^{\prime} + \sum_{i=k+1}^{W} (a_i^{\prime} \times c_i)
\end{equation}
\noindent where $a_i^{\prime}$ and $a_k^{\prime}$ are the new performance ratios with the consideration of the persistence overhead.

Our two performance goals are formulated as follows. We want to select a code region to meet the two goals.
\begin{equation}
\label{eq:goal2}
l_k < t_s \qquad 
\end{equation}
\begin{equation}
\label{eq:goal1}
Y^{\prime} > \tau   
\end{equation}
\textbf{Our algorithm to solve the problem.}
To determine if the selection of a code region can meet Equation~\ref{eq:goal2}, we need to estimate the performance loss ($l_k$) caused by persisting critical data objects. $l_k$ can be estimated based on measuring the overhead of flushing one cache block and the total number of cache blocks to flush. 
To determine if the selection of a code region can meet Equation~\ref{eq:goal1}, we use the following method to estimate $c_k^{\prime}$ (without doing extensive crash tests) and then calculate $Y^{\prime}$ based on Equation~\ref{eq:new_Y} and $c_k^{\prime}$. 

$c_k^{\prime}$ depends on how frequently we persist data objects in the code region. (1) If the code region $k$ is a loop structure, we can persist data objects at every iteration of the loop to maximize recomputability ($c_k^{max}$), or persist them every $x$ iterations 
with the recomputability $c_k^{x}$. If we do not persist them at all, then the recomputability of the code region is not changed (still $c_k$), and the code region is literally not selected. (2) If the code region is not a loop structure, we flush cache blocks at the end of the code region to reach $c_k^{max}$, or do not flush at all with no change of recomputability (still $c_k$).

To approximate $c_k^{max}$, we measure the recomputability of the code region $k$,  when persisting data objects at every code region and every iteration of the loop in each code region. We use the measured recomputability 
at the code region $k$ 
as $c_k^{max}$. To approximate $c_k^{x}$ for a code region with a loop structure, we use Equation~\ref{eq:approx_c_k_x}. 
\startcompact{small}
\begin{equation}
\label{eq:approx_c_k_x}
    c_k^{x} =  (c_k^{max} - c_k)\times \frac{1}{x} + c_k
\end{equation}
\stopcompact{small}
In essence, Equation~\ref{eq:approx_c_k_x} estimates $c_k^{x}$ based on a linear interpolation between $c_k^{max}$ and $c_k$. 

Using the above formulation and method, we can know the performance loss ($l_k$) and the application recomputability (
$c_k^{\prime}$
) for any code region, where $1 \le  k \le W$. 

Based on the above discussion, 
we can generalize our method to select multiple code regions as follows. To meet the two performance goals, we try to select those code regions with total performance loss less than $t_s$, and the recomputability after persisting data objects in the selected code regions should be larger than $\tau$. We also want to maximize recomputability. This is a variant of the 0-1 knapsack problem~\cite{knapsackbook} with the performance loss as the item weight and recomputability as the item value. This problem can be solved by the dynamic programming in pseudo-polynomial time.

\textbf{How to use the algorithm.}
To use the algorithm, we need to know the performance loss $l_k$ for each code region. The performance loss can be different 
with different data persisting frequencies. 
We estimate the performance loss based on the overhead measurement of flushing one cache block. Note that certain cache flushing instructions (\texttt{CLFLUSH} and \texttt{CLFLUSHOPT}) invalidate cache lines after cache flushing. This means that cache blocks need to be reloaded into the cache when they are re-accessed, which causes extra performance loss. To account for such performance loss, we double our estimation on the overhead of flushing cache blocks. 

To use the algorithm, we also need to know the recomputability of each code region without persisting any data object ($c_k$) and the recomputability of each code region after we persist critical data objects at the code region ($c_k^{\prime}$).
The two recomputability results can be measured by two campaigns of crash tests. In the first campaign, we do not persist any data object; In the second campaign, we persist critical data objects at every code region. For both campaigns, we measure the recomputability of each code region. 


\textbf{Discussions.}
When we estimate $l_k$, we assume every cache block of data objects is in the cache, which overestimates the performance overhead. Some cache blocks may not be in the cache and flushing them is not expensive. However, overestimation ensures that the real runtime overhead is smaller than $t_s$, which is good. 


To calculate $Y^{\prime}$, we use one campaign of crash tests to measure the recomputability of each code region, by persisting critical data objects at each code region. This method, although avoids massive numbers of crash tests, introduces the measurement inaccuracy, because persisting data objects in one code region can impact the recomputability of another code region. In essence, we ignore the possible propagation of computation inaccuracy from one code region to another. Such a method can make the measured recomputability smaller than the real recomputability. This means \name should result in larger recomputability and larger performance benefit in reality, which is good.

\subsection{Workflow of \name}
We present the whole workflow of \name in this section. The workflow includes four steps. 


Step 1: Running a crash test campaign. We collect the data inconsistent rate of candidates of critical data objects and calculate corresponding recomputability.


Step 2: Selection of data objects. We calculate the correlation between the inconsistent rate of data objects and application recomputability to decide critical data objects.


Step 3: Selection of code regions. We run another crash test campaign that persists critical data objects. The output of this step is the decision on where \name should flush critical data objects and how frequently that should happen.


Step 4: Production run. Just run the application, and \name automatically manages cache flushes.

The above steps are based on minor changes to the application, discussed as follows.

\textbf{Application preparation.} The application changes include two parts: (1) Allocating data objects that are referenced in the main computation loop and not read-only, with an \name API. Those data objects are candidates of critical data objects, and their addresses are passed into \name for potential cache flushing during production runs. (2) Identifying the end of first-level inner loops with an \name API. Those places delineate code regions. For (1) and (2), the compiler can annotate the application with the APIs, freeing the programmer from changing the application. For (1) with the pointer alias problem, the programmer has to manually allocate data objects, but the change is straightforward.

Among all benchmarks we evaluate, the change is less than 10 lines in each benchmark. Those benchmarks with changes can be found in our released code.


\section{Evaluation}
\label{sec:evaluation}

In this section, we study whether \name can effectively improve application recomputability and  
the runtime overhead of \name.
In the next section, we evaluate system efficiency of \name in large scale systems in the context of C/R mechanisms.  We use the benchmarks shown in Table~\ref{tab:bench}. 
Table~\ref{tab:hardware} shows the hardware we use for evaluation (including crash emulation and performance study).
To calculate application recomputability, we use the method in Section~\ref{sec:motivation_examples_setup} for crash tests. 
\begin{table}[t]
\caption{Hardware used in evaluation.}
 \label{tab:hardware}
\begin{tabular}{l}
\hline
CPU:  \phantom{LAL} Two Xeon Gold 6126 processors (Skylake) @ 2.6 GHz                      \\
DRAM:  \phantom{LA}    DDR4 129GB   BW: 106 GB/s  \phantom{L}  Latency: 87 ns \\ \hline
\end{tabular}
\end{table}

We set $t_s$ as 3\% in this section. 
We also use $t_s = 2\%$ and $5\%$ for the  \textit{sensitivity study}. In all tests, the runtime overhead is effectively bounded by $t_s$. But a smaller $t_s$ leads to less frequent persistence operations. As a result, a few benchmarks (e.g., FT) cannot meet the recomputation requirement imposed by $\tau$. We show the results of $t_s = 3\%$ in this section. We do not present the results for EP, because its inherent recomputability is 0. Even with \name, its recomputability is less than 3\%, and \name cannot bring benefit in system efficiency 
according to our model (Equation~\ref{eq:goal1}).

\begin{figure}[!t]
\centering
\includegraphics[width=0.50\textwidth]{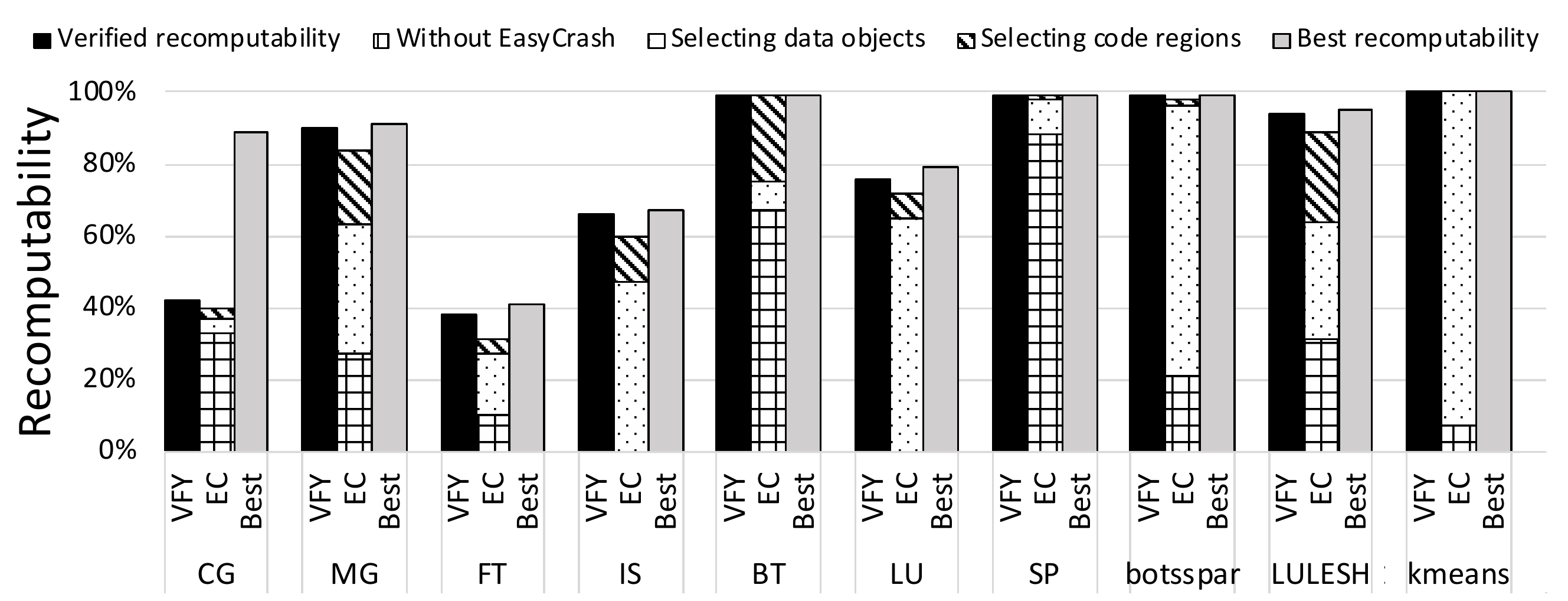}
\caption{
Application recomputability with different methods. Figure annotation: ``EC'', ``best'', and ``VFY'' stand for \name, best recomputability, and verified recomputability, respectively.}
\label{fig:recompute_success_rate_breakdown}
\end{figure}

\textbf{Effectiveness of \name.} Figure~\ref{fig:recompute_success_rate_breakdown} shows the application recomputability before and after we apply \name. To reveal the contributions of our two techniques (i.e., selecting data objects and selecting code regions) to improve the application recomputability, we first measure recomputability without using the two techniques (shown as ``without \name'' in the figure). Then we select data objects and persist them at the end of each iteration of the main computation loop. 
The recomputability improvement is shown as ``selecting data objects''.  We then select code regions to persist the selected data objects, and the recomputability improvement is shown as ``selecting code regions''.

To show the effectiveness of \name, we also show the \textit{best recomputability results}, and compare them with those after applying \name. We obtain the best recomputability by persisting critical data objects at each code region 
(if the code region has a loop structure, then we persist critical data objects at the end of each iteration of the loop). 
Note that the method to get the best recomputability is very costly (shown in the last column of Table~\ref{tab:exe_time}), which is not a practical solution for HPC. In Section~\ref{sec:design}, we have shown that persisting critical data objects can achieve very similar recomputability as persisting all data objects; hence, we do not show results for persisting all data objects.

We have the following observations from Figure~\ref{fig:recompute_success_rate_breakdown}. 
(1) \name achieves very high recomputability. Except for CG, the recomputability of applying \name is pretty close to the best one,
with a difference of only 5\% on average. 
For CG, there is a big difference (49\%), because many successful recomputation tests in CG require extra iterations, which is not acceptable in \name due to the concerns on runtime overhead. Note that even with the big difference, \name still brings 4\% improvement in system efficiency for CG (shown in Section~\ref{sec:checkpoint_emulation}). 

(2) \name significantly improves application recomputability. This fact is especially pronounced in the benchmarks MG, botsspar and kmeans. We see 56\%, 77\%, and 93\% improvement for the three benchmarks respectively. The average recomputability of all benchmarks after using \name is 82\%, while it is 28\% before using \name. Also, \name is able to transform 54\% of crashes that cannot correctly recompute into the correct computation.

\textbf{Result verification.} The above recomputation results are collecting using \tname. To verify the result correctness, we perform crash tests on a physical machine (see Table~\ref{tab:hardware}) without \tname. We randomly stop the application to emulate a crash and then make a copy of all candidates of critical data objects for restart. When the application restarts, we use the same method as in Section~\ref{sec:NVCT} (but no \tname). We measure application recomputability by crash test campaigns. Note that this method is different from a real crash, because when making the data copy, the system forces \textit{all} data objects to be consistent between caches and NVM. In a real crash and \tname, non-critical data objects are not consistent. Hence, this crash test should show stronger recomputability. Figure~\ref{fig:recompute_success_rate_breakdown} shows the results (see ``Verified''). As expected, the new result (recomputability) is higher than that of using \tname, but the two results are pretty close (less than 5\% difference), verifying the effectiveness of \name.

\begin{table}[h]
\scriptsize
\caption{Normalized execution time. ``Norm'' = ``normalized''. ``EC'' = ``\name''. ``best'' = ``the best recomputability''} \label{tab:exe_time}
\begin{tabular}{|P{0.8cm}|P{1.6cm}|P{1.2cm}|P{0.7cm}|P{0.8 cm}|P{1.1cm}|}
\hline
         & \begin{tabular}[c]{@{}l@{}}Time for persisting \\ critical data for once \end{tabular} & \begin{tabular}[c]{@{}l@{}}\# of persistence \\ operations \end{tabular} & \begin{tabular}[c]{@{}l@{}}Norm.\\exe. time \\ with EC \end{tabular} & \begin{tabular}[c]{@{}l@{}}Norm. \\exe. time \\ without EC \end{tabular} & \begin{tabular}[c]{@{}l@{}}Norm. exe. \\time achieving\\the best. \end{tabular}\\ \hline
\textbf{CG}       & \textless{}0.001 s   & 75   & 1.004    &   1.20  & 1.24\\ \hline
\textbf{MG}       & 0.035 s & 40   & 1.012 &   1.26 & 1.24\\ \hline
\textbf{FT}       & 0.032 s & 80       & 1.016     &  1.22 & 1.12\\ \hline
\textbf{IS}       & 0.030 s   & 10   & 1.011  & 1.15 & 1.43\\ \hline
\textbf{BT}       & 0.034 s  & 200    & 1.025   & 1.10 & 1.34\\ \hline
\textbf{SP}       & 0.034 s   & 200   & 1.022  & 1.23 & 1.55\\ \hline
\textbf{LU}       & 0.033 s & 250 & 1.025 & 1.23 & 1.58\\ \hline
\textbf{botsspar} & 0.030 s & 200   & 1.015   &1.28 & 1.62\\ \hline
\textbf{LULESH}   & 0.030 s & 293 &  1.016  & 1.25 & 1.43\\ \hline
\textbf{kmeans}   & \textless{}0.001 s    & 36  & 1.000  &1.00 & 1.00\\ \hline
\textbf{Average}   & about 0.026 s    & 138  & 1.015 &1.19 & 1.35\\ \hline
\end{tabular}
\end{table}

\begin{figure*}
\centering
\includegraphics[width=1.0\textwidth]{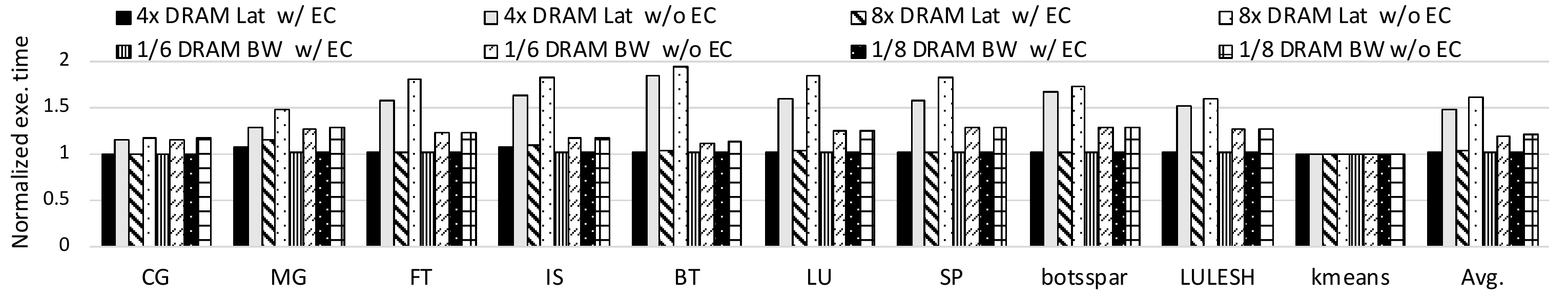}
\caption{
The performance (normalized execution time) with and without \name. Figure annotation: ``EC'' stands for \name; ``Lat'' stands for latency; ``BW'' stands for bandwidth.} 
\label{fig:sensitivity}
\end{figure*}

\textbf{Performance overhead.} 
We measure the runtime overhead of persisting critical data objects at critical code regions with \name, with no crash triggered. We emulate NVM with DRAM, and leverage \textit{CLFLUSHOPT}~\cite{inteloptimize} for best performance of cache flushing.
Table~\ref{tab:exe_time} shows the results. 
The table reports the execution time of persisting critical data objects for once (i.e., \textit{one persistence operation}), the number of persistence operations with \name, and total execution time with persistence operations. 
\textit{In the rest of this section, the total execution time is normalized by the execution time without any persistence operation.}
In general, the runtime overhead is no larger than 2.5\% (bounded by $t_s=3\%$). 
For comparison purpose, we show the overhead of persisting all candidate data objects at the end of each iteration of the main computation loop (shown in the fifth column of the table), which is a case without the selection of data objects and code regions. This case causes a high overhead (19\% on average), much larger than \name. We also evaluate the overhead of achieving the best recomputability by persisting critical data objects at the end of each 
code region. 
The runtime overhead is 35\% on average, which is much larger than \name.

\textbf{Performance study with various NVM bandwidth/latency.}
We study the performance of \name using a DRAM-based emulator, Quartz~\cite{middleware15:volos}. With Quartz, we emulate NVM configured with 4x and 8x DRAM latency, or 1/6 and 1/8 DRAM bandwidth. Such NVM configurations correspond to the performance of Optane PMM and PCM~\cite{pcm_book}. 
Figure~\ref{fig:sensitivity} shows the result. \name brings less than 9\% (2.3\% on average) runtime overhead in all case.

We further compare the performance of \name and no \name (without \name, we persist all candidate data objects at the end of each iteration of the main computation loop). Figure~\ref{fig:sensitivity} shows the result. Without \name, the performance overhead is 48\%, 62\%, 21\% and 22\% on average for the configurations of 4x and 8x DRAM latency, and 1/6 and 1/8 DRAM bandwidth respectively. We conclude that using \name performs better than no use of \name on NVM with various performance.

\begin{figure}[!t]
\centering
\includegraphics[width=0.48\textwidth]{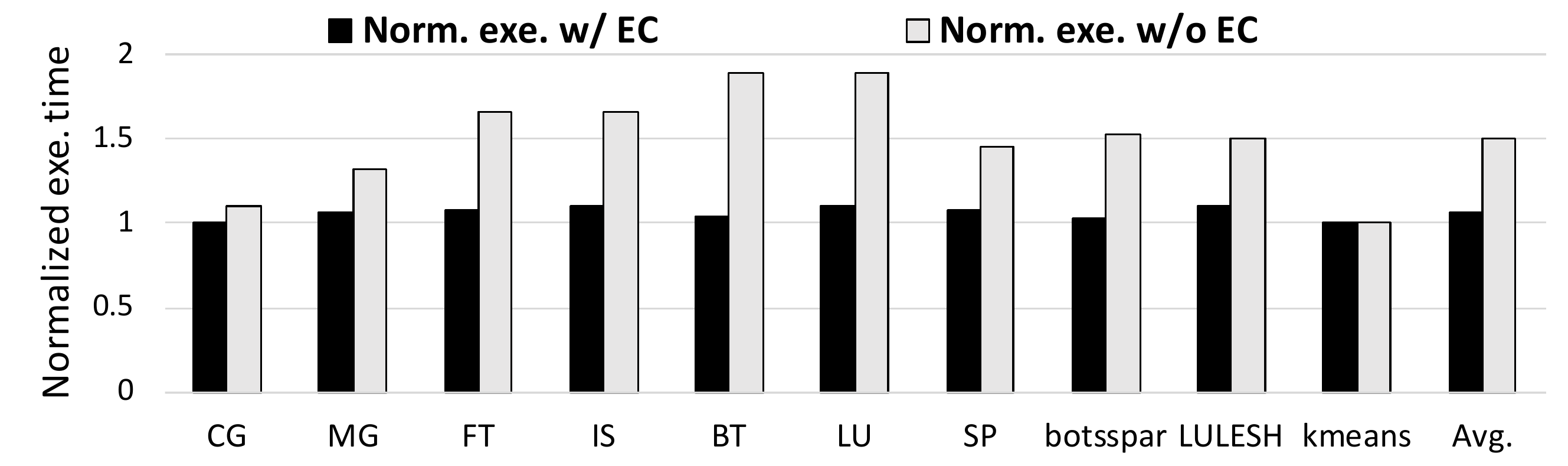}
\caption{
The performance(normalized execution time) with and without EasyCrash on Intel Optane PC DMM. 
}
\label{fig:optane}
\end{figure}

\textbf{Performance study with Optane DC PMM.}
We further evaluate the performance of \name on the very recent Intel Optane DC PMM, shown in Figure~\ref{fig:optane}. \name incurs only 6\% performance overhead on average, while without using \name the performance overhead is 50\% on average.

\textbf{Reduction on the number of writes.}
NVM has limited write endurance. It is critical for NVM to avoid additional writes. We compare \name and the traditional C/R mechanism in terms of the number of extra writes in NVM. For \name, the extra writes come from persisting critical data objects at critical code regions by using cache flushing instructions. As discussed in Section~\ref{sec:cflash_data_persist}, when cache blocks of critical data objects are clean or not resident in the cache, flushing them does not cause any write in NVM. For the traditional C/R mechanism, the extra writes come from making a copy of data objects; the extra writes also come from cache line eviction because of loading checkpoint data into the cache when making data copy~\cite{lazypersistency:isca18}. We use \tname to measure the number of writes in NVM. Whenever a dirty cache block is written back from the last level cache to NVM, we count the number of writes by one. 

To enable a fair comparison with \name, we perform the C/R in two ways: in one way, we checkpoint critical data objects, and in the other way we checkpoint all data objects (not including read-only ones).
Also, we assume that checkpoint happens only once. This is a rather conservative assumption. The checkpoint could happen more often (depending on the system failure rate and application execution time), which causes the more extra number of writes. We consider the system failure rate and application execution time to evaluate the effects of checkpoint in Section~\ref{sec:checkpoint_emulation}. 

\begin{figure}[!t]
\centering
\includegraphics[width=0.48\textwidth]{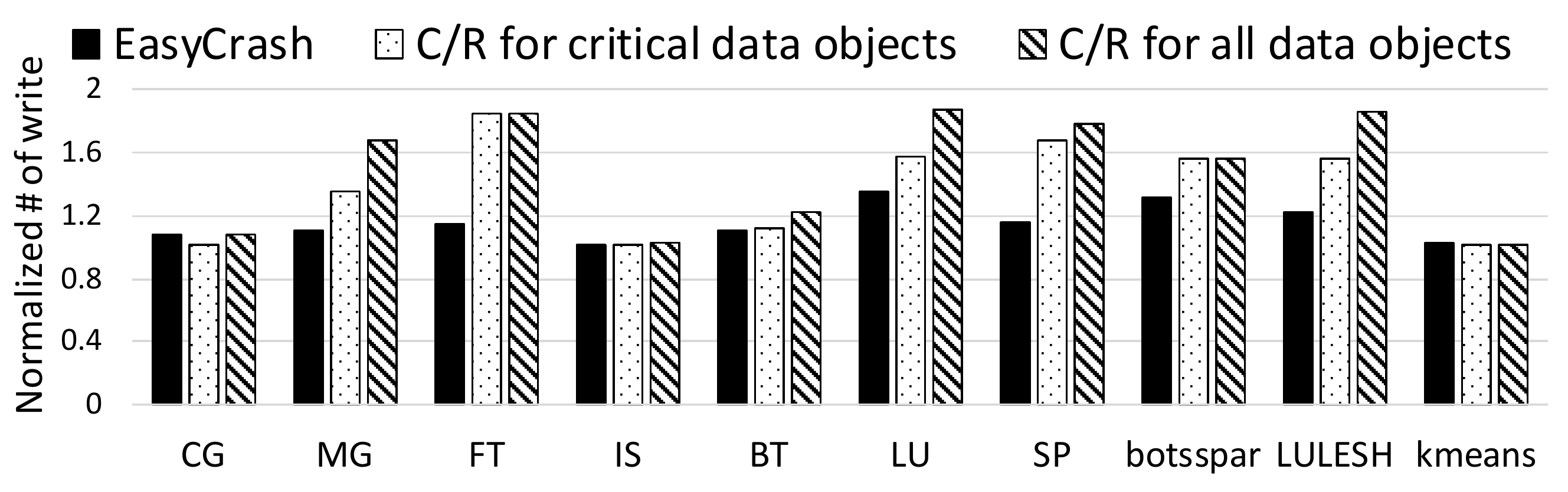}
\caption{
Normalized number of NVM write. }
\label{fig:extra_nvm_write}
\end{figure}

Figure~\ref{fig:extra_nvm_write} shows the number of NVM write normalized by the total numbers of writes in NVM without \name and C/R. 
On average, \name adds 16\% additional writes, while C/R adds 38\% and 50\% for the two methods of checkpointing respectively. Also, for those benchmarks with large data objects (e.g., FT, SP, and LU), \name is especially beneficial for reducing the number of writes. This is because the number of extra writes for persisting data objects in a persistence operation is bounded by the number of cache lines in the last level cache. A larger data object indicates that \name flushes more clean cache lines or non-resident cache blocks without causing actual writes in NVM. For benchmarks with small data objects (e.g., CG), \name is not beneficial to reduce the number of writes, but writing those small data objects in NVM do not usually cause the serious endurance problem.

\section{End-to-End Evaluation}
\label{sec:checkpoint_emulation}
We evaluate \name in the context of large-scale parallel systems running time-consuming HPC applications with a C/R mechanism. To enable convincing evaluation, we need different system scales with various configurations, which is expensive to achieve. We develop an emulator based on performance models and performance analysis in Section~\ref{sec:evaluation}. 
Table~\ref{tab:annotation} summarizes the model annotation.

\textbf{Basic assumptions. } 
We assume that 
the checkpointing process does not have any corruption. This is a common assumption~\cite{Bosilca:2014:UMA:2768028.2768033, hpdc17:fang}. We model synchronous coordinated checkpointing, which is the most common C/R in HPC. With this mechanism, all nodes take checkpoints at the same time with synchronization. 
The checkpoints are saved in local storage and then \textit{asynchronously} moved to remote storage nodes. This checkpoint mechanism is the most common practice in HPC and commonly used in the recent existing work~\cite{hpdc17:fang,DBLP:journals/tpds/MohrorMBS14, 16-moody2010design}.
When a crash happens in one node and the application cannot successfully run to the completion or pass the acceptance verification after restarting using \name, all nodes will go back to the last checkpoint. Note that with \name, the application has a high probability to successfully recompute after the restart. 
Hence, the checkpoint interval with \name is longer. 

\textbf{Performance emulation. }
Our emulator includes system and application related parameters. We summarize the \textit{system related parameters} as follows.

\begin{enumerate}[leftmargin=*]
\item \textbf{\textit{MTBF}}: Mean time between failures of the system without EasyCrash. $MTBF_{EasyCrash}$ is \textit{MTBF} with \name. Since the average application recomputability with \name is 82\% (Section~\ref{sec:evaluation}),we have 
$MTBF_{EasyCrash} = {MTBF}/(1-82\%)$.
   
\item \textbf{\textit{T\_chk}}: The time for writing a system checkpoint. The checkpoint on each node is written into local SSD (not in NVM main memory) and then gradually migrated to remote storage nodes (the data migration overhead is not included in \textit{T\_chk}). Such a multi-level checkpoint mechanism is based on~\cite{DBLP:journals/tpds/MohrorMBS14}. 
The checkpoint data should not be written into NVM-based main memory, 
because it significantly reduces memory space useful for applications. 

\item \textbf{\textit{T\_r}}: The time for recovering from the previous checkpoint. Similar to the existing work~\cite{Bosilca:2014:UMA:2768028.2768033}, we assume \textit{T\_r} = \textit{T\_chk}. 

\item \textbf{\textit{T\_sync}}: The time for synchronization across nodes. We use the assumption in~\cite{Fang:2017:LLC:3078597.3078609}: The synchronization overhead is a constant value, and we use 50\% of the checkpoint overhead as \textit{T\_sync}. 

\item \textbf{\textit{T}}: The checkpoint interval, based on Young's formula~\cite{Young:1974:FOA:361147.361115}, $T = \sqrt{2 \times \textit{T\_chk} \times MTBF}$. This formula has been shown to achieve almost identical performance as in realistic scenarios~\cite{6968777}. 

\item \textbf{\textit{T\_vain}}: The wasted computation time. When the application rolls back to the last checkpoint, the computation already performed in the checkpoint interval is lost. 
On average, half of a checkpoint interval for computation is wasted (i.e., $T\_vain = 50\% \times T$).
\end{enumerate}

We summarize the \textit{application related parameters} as follows.
\begin{enumerate}[leftmargin=*]
\item $R_{EasyCrash}$: The application recomputatbility achieved by using \name.  
\item $t_s$: The runtime overhead introduced by \name because of persisting critical data objects (e.g., 3\% in our evaluation). 
\end{enumerate}
\vspace{-3pt}



Based on the above notations, we use performance models to evaluate system efficiency. 
The system efficiency is the ratio of the accumulated useful computation time ($u$) to total time spent on the system ($Total\_Time$), which is ($u/Total\_Time$). 
We assume that the accumulated useful computation takes checkpoints $N$ times; and during the whole computation, the crash happens $M$ times. 

Equation~\ref{eq:cost} models the total time spent on the HPC system without using \name. The equation includes useful computation and checkpoint time ($N \times (T+T\_chk)$), and the cost of recovery using the last checkpoint ($M \times (T\_vain+ T\_r + T\_sync)$). The number of crashes ($M$) is estimated using Equation~\ref{eq:crash}.
\begin{equation}
\label{eq:cost}
Total\_Time = N \times(T + T\_chk) + M \times (T\_vain+ T\_r + T\_sync)
\end{equation}
\begin{equation}
\label{eq:crash}
M = \frac{Total\_Time}{MTBF}
\end{equation}

\name improves HPC system efficiency by avoiding large recovery cost from the last checkpoint and increasing the checkpoint interval. \name also brings ignorable runtime overhead. 
Equation~\ref{eq:easycrash_cost} models the total execution time with \name ($Total\_Time^\prime$), where $N'$ and $T'$ are the number of checkpoints and their interval when using \name, and $M'$ is the number of crashes that have to go to the last checkpoint for recovery, and $M''$ is the number of crashes that use \name to recompute successfully. 
\begin{equation} \label{eq:easycrash_cost}
\begin{split}
Total\_Time^\prime = & \, N^\prime \times (T^\prime + T\_chk) \, + \\ 
                          & M^\prime \times (T\_vain'+ T\_r + T\_sync) \, + \\
                          & M'' \times (T\_r^\prime + T\_sync)
\end{split}
\end{equation}
\begin{equation} \label{eq:new_M}
 M' = M \times (1 -R_{EasyCrash}), \qquad M'' = M \times R_{EasyCrash}
 \end{equation}
With \name, the checkpoint interval ($T^\prime$) becomes longer ($T^\prime > T$), and also should include a small runtime overhead ($t_s$). 
As a result, the number of checkpoints ($N^\prime$) becomes smaller ($N^\prime < N$), and the checkpoint overhead ($N^\prime \times T_{chk}$) becomes smaller. 
With and without \name, the useful computation remains similar because of small runtime overhead of \name (i.e., $N^\prime \times T^\prime \approx N \times T$). 
To calculate $T^\prime$, 
we use Young's formula, 
$T^\prime = \sqrt{2 \times \textit{T\_chk} \times MTBF_{EasyCrash}}$.

With \name, once a crash happens, the system either goes to the last checkpoint with recovery overhead modeled as ($M^\prime \times (T\_vain+ T\_r + T\_sync)$), or uses \name to restart and successfully recompute with recovery overhead modeled as ($M'' \times (T\_r^\prime + T\_sync)$). With NVM and \name, the recovery cost $T\_r$ becomes $T_r^\prime$, which becomes smaller, because we load data objects from NVM-based main memory, not from local SSD or storage node. $T_r^\prime$ is estimated using the total data size of non-readonly data objects divided by NVM bandwidth (DRAM bandwidth in our evaluation). 



\textbf{Choice of parameters.} 
The time spent on writing a checkpoint to persistent storage depends on hardware characteristics. A modern HPC node normally has 64 to 128 GB memory. For nodes using SSD and NVMe, the average I/O bandwidth is 2GB/s; For nodes using HDD, the average I/O bandwidth is around 20 MB/s to 200 MB/s~\cite{Bhimji2016AcceleratingSW,wu2017early}. As a result, we choose the checkpointing overhead ($T\_chk$) as 32s, 320s, 3200s to represent different hardware scenarios. A similar set of values is used in previous efforts~\cite{Fang:2017:LLC:3078597.3078609, Bosilca:2014:UMA:2768028.2768033,6968777}. 
We emulate the system with 100,000 nodes for a long simulation time (10 years, i.e., $Total\_Time$ and $Total\_Time'$ are 10 years). Previous work~\cite{6903615} shows that systems in such a scale usually experience around 2 failures per day ($MTBF$ = 12 hours). Based on this data, we scale $MTBF$  as in~\cite{Fang:2017:LLC:3078597.3078609} for 200,000 and 400,000 nodes. As a result, $MTBF$ for them are 6 and 3 hours respectively. 
\begin{figure}[!t]
\centering
\includegraphics[width=0.48\textwidth]{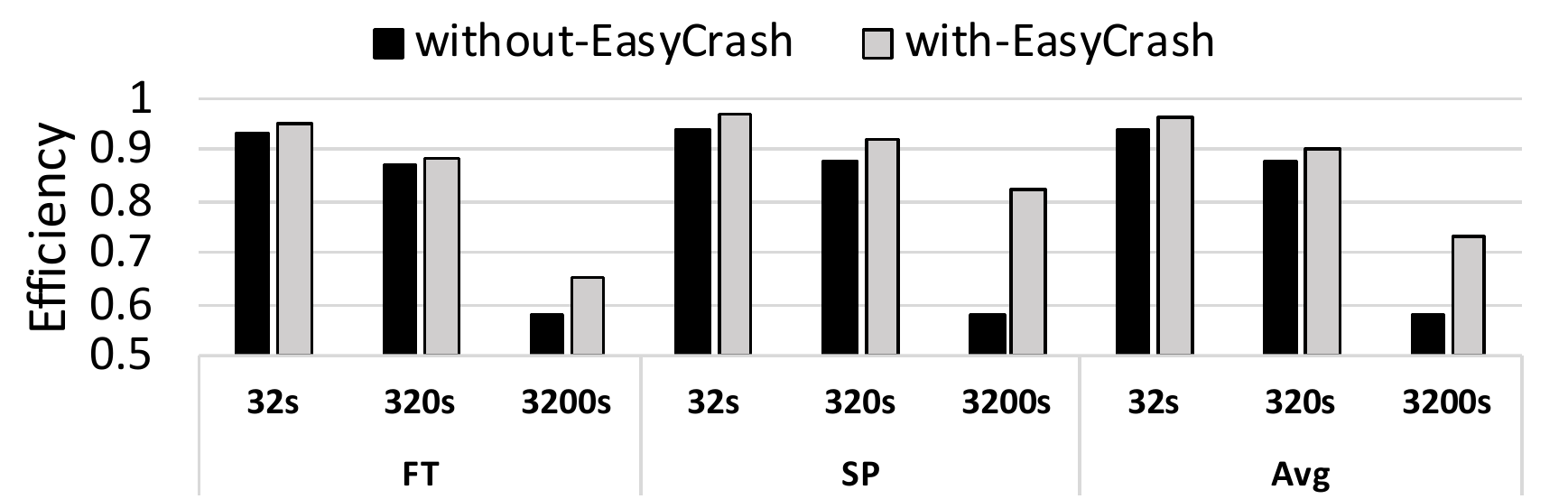}
\caption{System efficiency without and with \name when the system MTBF is 12 hours. The x-axis shows different checkpointing overhead. ``Avg'' stands for ``average''.}
\label{fig:efficiency}
\end{figure}

\textbf{Results for system efficiency.}
Figure~\ref{fig:efficiency} shows the system efficiency with and without \name under different checkpoint overhead. Because of the space limitation, we show the benchmarks with the lowest (FT) and highest recomputability (SP). \textit{We also show the average values of \textbf{all benchmarks}}. 
\name improves system efficiency by 2\%-24\%. On average, the system efficiency with \name is improved by 2\%, 3\% and 15\% when the checkpoint overhead is 32s, 320s, and 3200s respectively.
\begin{figure}[!tbp]
\centering
    \begin{subfigure}[b]{0.48\linewidth}
        \includegraphics[width=\textwidth]{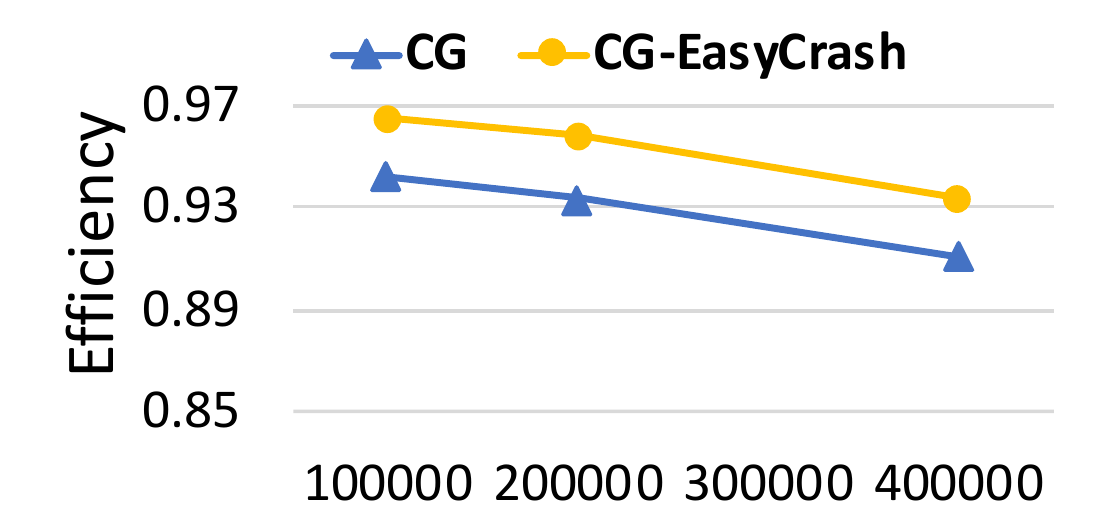}
        \subcaption[a]{$T\_chk=32s$.}
    \label{fig:se_32}
        \end{subfigure}
    \begin{subfigure}[b]{0.48\linewidth}
    \includegraphics[width=\textwidth]{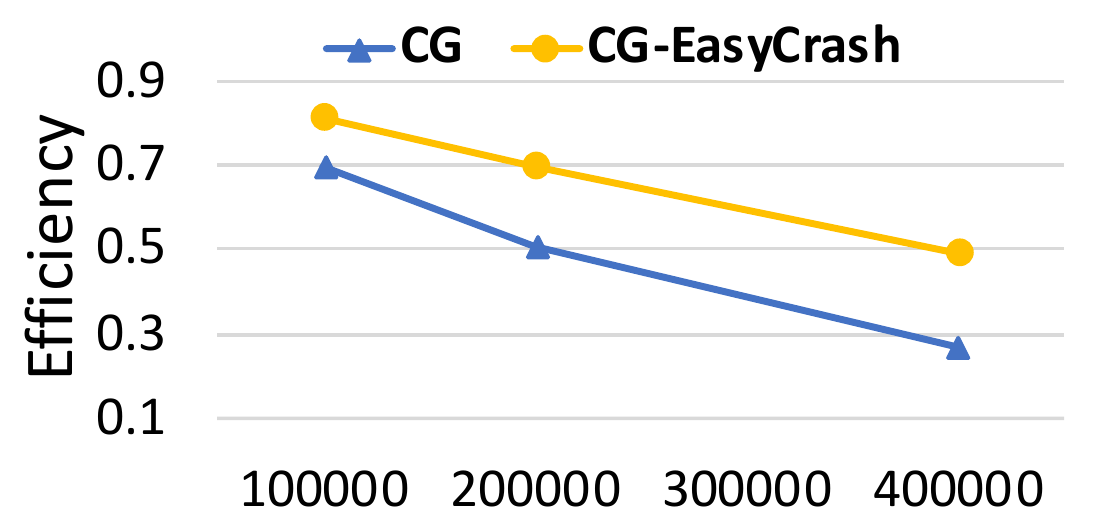}
            \subcaption[b]{$T\_chk=3200s$.}
        \label{fig:se_3200}
        \end{subfigure}
	    \caption{System efficiency for CG without and with \name when the system scales from 100,000 to 200,000 and 400,000 nodes} 
\label{fig:efficiency_scale}
\end{figure}

Furthermore, we evaluate the system scalability with \name. We evaluate all benchmarks but only present CG because of space limitation. Figure~\ref{fig:efficiency_scale} shows the efficiency with and without \name in different system scales. 
With \name the system efficiency always outperform that without \name.
\textit{This trend is consistent with \textbf{all benchmarks}}. The system with \name achieve better efficiency as the system scale increases.


\textbf{Results for the number of writes.} 
With \name, the number of \textit{additional writes} is reduced by 32\% to 57\% (44\% on average), comparing with the traditional C/R without \name. 

\textbf{Determination of recomputability threshold $\tau$}. To ensure the system with \name has higher efficiency than with C/R (i.e., $u/Total\_Time' > u/Total\_Time$), the application recomputability  must be higher than a threshold $\tau$ (see Section~\ref{sec:selection_code_region}).  This means $Total\_Time' < Total\_Time$. Given $Total\_Time$ and Equations~\ref{eq:easycrash_cost} and~\ref{eq:new_M}, we can calculate a lower bound for $R_{EasyCrash}$, which is $\tau$.
\section{Discussions}
\label{sec:discussion}
\textbf{Determining how/when to use \name. }
To decide whether to use \name, we need to have multiple information, including (1) system MTBF, (2) the checkpoint overhead, (3) the application recomputability with \name to select data objects and code regions and estimate efficiency benefit, and (4) the acceptable minimum performance loss $t_s$. For (1), (2) and (4), it is reasonable to assume that the system operator has such information. With (1), (2) and (4), the recomputability threshold $\tau$ can be calculated. 
For (3), we use crash tests, but we can avoid them by an application characterization study. In particular, we can detect computation patterns that tolerate computation inaccuracy as in~\cite{sc18:guo}. Then we set up a model to correlate those patterns and application recomputability. Given an application, we simply count those patterns and use the model to predict recomputability without any crash test.
\textbf{What kind of application is not suitable?}
We found that there are two categories of applications not suitable for \name. (1) Applications with small data objects and small memory footprint. When a crash happens to the application, most of the application data are resident in the cache and lost. To ensure high recomputability, we have to persist data objects frequently, which causes high runtime overhead. (2) Applications with no tolerance for computation errors. These applications regard any application outcome different from that of the golden run as incorrect. Many of our crash-and-restart tests generate outcomes different from those of the golden run, but these tests pass the acceptance verification.  

For (1), the system can disable \name and only employ the traditional checkpoint mechanism to handle failures. Because of the small memory footprint of the application, the checkpoint is small and can be stored in NVM with small overhead.

For (2), when the application outcome is different from that of the golden run, the users can claim a silent data corruption (SDC) happens~\cite{Kai:icpp18,sc18:guo}. With the acceptance verification, many applications treat this kind of SDC as benign and ignorable. Examples of these applications include many iterative solvers and machine learning training workloads, which have been leveraged in the recent approximate computing research~\cite{ics06:rinard, 19-mengte2010exploiting, oopsla13:carbin, oopsla14:misailovic, onward10:rinard, fse11:douskos}. The applications that cannot tolerate SDC cannot use \name.



\section{Related Work}
Some research efforts focus on establishing crash consistency in NVM~\cite{Dulloor:eurosys14, nvheap:asplos11, Kolli:ASPLOS2016, Mnemosyne:ASPLOS11, cdds:fast11} by software- and hardware-based techniques. Building an atomic and durable transaction by undo logging and redo logging mechanisms in NVM is the most common method to enforce crash consistency. Intel Persistent Memory Development Kit (PMDK)~\cite{pmdk} adopts an undo logging mechanism that keeps an unmodified log copy before any in-place update in NVM happens. Volos et al.~\cite{Mnemosyne:ASPLOS11} use a redo mechanism for programming in persistent memory. 
Some work on NVM-aware data structures~\cite{nvtree:fast15,cdds:fast11} re-design specific data structures to explicitly trigger cache flushing for crash consistency. However, the existing work can impose big performance overhead and extensive changes to the applications, which may not be acceptable by HPC. Different from the above work that relies on strong guarantees on crash consistency and heavily involves programmers to enforce crash consistency, \name enables automatic exploration of application recomputability without extensive changes to applications.


A few recent efforts focus on using NVM for HPC fault tolerance~\cite{loopbased:pact17, shuo:cluster17, lazypersistency:isca18}. They avoid flushing caches for high performance, and rely on algorithm knowledge~\cite{shuo:cluster17} or high requirements on loop structures~\cite{loopbased:pact17, lazypersistency:isca18} to recover computation upon application failures. \name is significantly different from them: \name aims to explore application's intrinsic fault tolerance and leverage consistent \textit{and} inconsistent data objects for recomputation;
\name is general, because it does not have high requirement on code structure or application algorithms. 


Many existing works focus on approximate computing, and trade computation accuracy for better performance by leveraging application intrinsic fault tolerance~\cite{pldi11:sampson, oopsla14:misailovic, asplos14:samadi}. \name, in essence, belongs to approximate computing, but applies the idea to NVM and HPC, which is unprecedented. LetGo~\cite{Fang:2017:LLC:3078597.3078609} is another example of approximate computing. Once a failure happens, LetGo attempts to continue application execution. \name is significantly different from LetGo. \name loses dirty data in caches when a crash happens, and hence selectively flushes data objects in some code regions to ensure the improvement of system efficiency. Letgo does not lose data in caches and provides no guarantee on the improvement. LetGo does not consider differences of code regions and data objects in their impacts on application recomputability. \name is highly NVM oriented, while LetGo is not.

\section{Conclusions}
Large-scale HPC systems face a grand challenge on system reliability. The emerging NVM provides a new solution to address this challenge: Leveraging the non-volatility of NVM as main memory, we can retain data in NVM instead of losing them as in DRAM when a crash happens, and restart the HPC application using the retained data. This paper is the first one that studies the feasibility of the above solution and provides a comprehensive analysis on application recomputability. We provide a set of techniques to improve application recomputability and make the solution feasible and beneficial. We demonstrate large improvement in system efficiency with ignorable runtime overhead, and greatly reduce the number of writes for better NVM endurance. 


\clearpage

\bibliographystyle{ACM-Reference-Format}
\bibliography{kai,li,jie}
\clearpage



\end{document}